\documentclass[aps,twocolumn,showpacs,superscriptaddress,amsmath,amssymb]{revtex4}
\usepackage{graphics,epsf,colordvi}

\begin{document}

\title{Partial-wave analysis of multiphoton ionization of sodium by
femtosecond laser pulses of 800 nm wavelength in over-the-barrier
ionization regime}

\author{A. Bunjac}
\affiliation{Institute of Physics, University of Belgrade, P.O.
Box 57, 11001 Belgrade, Serbia}
\author{D. B. Popovi\'c}
\affiliation{Institute of Physics, University of Belgrade, P.O.
Box 57, 11001 Belgrade, Serbia}
\author{N. S. Simonovi\'c}
\affiliation{Institute of Physics, University of Belgrade, P.O.
Box 57, 11001 Belgrade, Serbia}

\begin{abstract}
Multiphoton ionization of sodium by laser pulses of 800\,nm
wavelength and 57\,fs duration is studied in the range of laser
peak intensities belonging to over-the-barrier ionization regime.
Photoelectron momentum distributions (PMD) and the energy spectra
are determined numerically by solving the time dependent
Schr\"odinger equation. The calculated spectra agree well with the
spectra obtained experimentally by Hart~{\em et~al.}~[Phys.~Rev.~A
\textbf{93}, 063426 (2016)]. The contributions of photoelectrons
with different values of the orbital quantum number in the PMD are
determined by expanding the photoelectron wave function in terms
of partial waves. Partial wave analysis of the spectral peaks
related to Freeman resonances has shown that each peak has
photoelectron contributions from different ionization channels
which are characterized by different photoelectron energies and
different symmetries of released photoelectron wave-packets. These
findings are justified by calculating the populations of excited
states during the pulse. Our analysis indicates that the
contribution of specific ionization channels in the total
photoelectron yield might be selectively increased by varying to
some extent the values of pulse parameters used here.
\end{abstract}

 \today

\maketitle

\section{Introduction}
\label{intro}

Strong-field ionization of the alkali-metal atoms has been studied
intensively over the past ten years, both experimentally and
theoretically including {\em ab initio} numerical calculations
\cite{wollenhaupt,krug,schuricke,JJ,morishita,schuricke2,hart2016,pccp,wessels}.
A specific feature of this group of atoms -- a low ionization
potential, which ranges from  $I_p \approx 3.89$\,eV (for cesium)
to 5.39\,eV (for lithium), causes that a considerably smaller
number of photons of a given energy $\hbar\omega$ is required for
their photoionization than for the ionization of other atoms. For
example, with the laser wavelength of around 800 nm ($\hbar \omega
\approx 1.55$\,eV) it takes four photons to ionize an alkali-metal
atom, unlike to the case of frequently used noble gases where this
number is of the order of ten. Since for a dipole transition
requiring $N$ photons the lowest order perturbation theory
predicts that the photon absorption rate $W$ is proportional to
the $N$-th power of the laser intensity $I$ ($W \sim I^N$ if $I
\ll I_a$, where $I_a = 3.50945 \times 10^{16}\,\mathrm{W/cm}^2$ is
the atomic unit value for intensity, see e.g. \cite{JKP}.),
measurable effects in experiments with multiphoton ionization
(MPI) of alkali can be observed at relatively low laser
intensities, available in table-top laser systems.

The perturbative treatment, however, is not applicable at higher
intensities which can be achieved today. One indication of the
nonperturbative regime is the so-called above threshold ionization
(ATI) \cite{mittleman,dk2000,JKP} in which the atom absorbs more
photons than the minimum required. Under these conditions the
photoelectron spectra (PES, electron yield versus their excess
energy $\epsilon$) were seen to consist of several peaks,
separated by the photon energy $\hbar\omega$, and appearing at
energies $\epsilon^{(s)} = (N_0 + s)\hbar\omega - I_p$, where
$N_0$ is the minimum number of photons needed to exceed the
ionization potential $I_p$ and $s = 0,1,\ldots$ is the number of
excess ("above-threshold") photons absorbed by the atom. (For the
alkali-metal atoms and the laser of 800\,nm wavelength, $N_0 =
4$.) By increasing the intensity over a certain value, $W$ does
not follow further the prediction $I^{N_0+s}$ of the perturbation
theory.

At even larger intensities, the electric component of the laser
field becomes comparable with the atomic potential, opening up
another ionization mechanism -- the tunnel ionization. In this
case the field distorts the atomic potential forming a potential
barrier through which the electron can tunnel. Multiphoton and
tunneling ionization regimes are distinguished by the value of
Keldysh parameter \cite{keldysh} which can be written as $\gamma =
\sqrt{I_p/(2U_p)}$, where $U_p = e^2F^2/(4m_e\omega^2)$ is the
ponderomotive potential of ejected electron with mass $m_e$ and
charge $e$. The value of the electric field $F$ in the expression
for $\gamma$ corresponds to the peak value of laser intensity.
Multiphoton and tunneling regimes are characterized by $\gamma \gg
1$ (high-intensity, long-wavelength limit) and $\gamma \ll 1$
(low-intensity, short-wavelength limit), respectively. The
transition regime at $\gamma \approx 1$ for alkali-metal atoms is
reached at considerably lower intensities than for other atoms,
again due to the small ionization potential $I_p$. The experiments
accessing the strong-field regime with alkali
\cite{schuricke,schuricke2,hart2016,wessels} have revealed that
the commonly used strong-field ionization models in the form of a
pure MPI or tunnel ionization cannot be strictly applied. The
problem, however, goes beyond by using an {\em ab-initio}
numerical method for solving the time-dependent Schr\"odinger
equation (TDSE).

Finally, at a sufficiently high laser intensity, the field
strength overcomes the atomic potential. This can be considered as
the limiting case of tunnel ionization when the barrier is
suppressed below the energy of atomic state. This regime is
usually referred to as over-the-barrier ionization (OBI). Such a
barrier suppression takes place independently of the value of
Keldysh parameter. For neutral atoms the threshold value of field
strength for OBI is estimated as $F_\mathrm{OBI} \approx I_p^2/4$
(in atomic units). $F_\mathrm{OBI}$ values for alkali, determined
more accurately, are given in Ref.~\cite{MS}. The corresponding
laser intensities can be obtained by formula $I = I_a F^2$, where
$F$ is expressed in atomic units and $I_a$ is the above introduced
atomic unit for intensity. For noble gas atoms irradiated by the
laser of wavelength from the visible light domain, OBI was
occurring well into the tunneling regime \cite{mevel}. This is,
however, not a general rule. For atoms with low ionization
potentials, as the alkali-metal atoms are, the OBI threshold,
compared to that for hydrogen or noble gases, is shifted to
significantly lower values of the field strength. For example, the
laser peak intensity that corresponds to the OBI threshold for
sodium is about $3.3 \, \mathrm{TW/cm}^2$ ($F_\mathrm{OBI} =
0.0097$\,a.u. \cite{MS}), whereas the value of Keldish parameter
for the sodium atom interacting with the radiation of this
intensity and 800\,nm wavelength is $\gamma = 3.61$. Thus, the OBI
threshold in this case belongs to the MPI regime. Previous
experiments and theoretical studies have already mentioned this
peculiar situation for sodium and other alkali
\cite{schuricke,schuricke2,hart2016,morishita,JJ,wessels}. In
addition, it is demonstrated that at intensities above the OBI
threshold the atomic target is severely ionized before the laser's
peak intensity is reached \cite{morishita}. Thus, the ionization
occurs at the leading edge of the pulse only, that is equivalent
to the ionization by a shorter pulse.

A remarkable feature of the photoelectron spectra obtained using
short (sub-picosecond) laser pulses is the existence of
substructures in ATI peaks, known as Freeman resonances. The
mechanism which is responsible for occurrence of these
substructures is the dynamic (or AC) Stark shift
\cite{mittleman,dk2000,dk1999} which brings the atomic energy
levels into resonance with an integer multiple of the photon
energy. Freeman et al.~\cite{freeman,gibons} have shown that when
atomic states during the laser pulse transiently shift into
resonance, the resonantly enhanced multiphoton ionization (REMPI)
\cite{dk2000,grossmann,JKP}) takes place, increasing the
photoelectron yield, and one observes peaks at the corresponding
values of photoelectron energy. Thus, the peaks in the PES can be
related to the REMPI occurring via different intermediate states.

The resonant dynamic Stark shift of energy levels corresponding to
sodium excited states $nl$ ($n \le 6$), relative to its ground
state (3s) energy, is recently calculated for the laser
intensities up to $7.9\,\mathrm{TW/cm}^2$ and wavelengths in the
range from 455.6 to 1139\,nm \cite{pccp}. These data are used to
predict the positions of REMPI peaks in the PES of sodium
interacting with an 800\,nm laser pulse. Freeman resonances in the
PES of alkali-metal atoms have been studied in papers
\cite{wollenhaupt,krug,schuricke,JJ,morishita,schuricke2,hart2016,pccp,wessels},
mentioned at the beginning of Introduction, where a number of
significant results have been reported.

The dynamic Stark shift also appears as an important mechanism in
the strong-field quantum control of various atomic and molecular
processes \cite{rabitz,shapiro,sussman,g-vazquez}. Focusing on the
MPI of atoms, a particular challenge would be the selective
ionization of an atom through a single energy level which could
produce a high ion yield. By increasing simply the laser intensity
one increases the yield, but also spreads the electron population
over multiple energy levels \cite{gibons} and, in turn, reduces
the selectivity. Krug {\em et al} \cite{krug} have shown in the
case of multiphoton ionization of sodium that chirped pulses can
be an efficient tool in strong-field quantum control of multiple
states. Hart {\em et al} in their paper \cite{hart2016} claim that
improved selectivity and yield could be achieved by controlling
the resonant dynamic Stark shift via intensity of the laser pulse
of an appropriate wavelength ($\sim 800$\,nm).

In this paper we study the photoionization of sodium by the laser
pulse of 800\,nm wavelength and 57\,fs duration with the peak
intensities ranging from $3.5$ to 8.8\,TW/cm$^2$, which belong to
OBI domain in the MPI regime and which have been used in the
experiment by Hart {\em et. al.} \cite{hart2016}. Using the
single-active-electron approximation we calculate the
corresponding photoelectron momentum distribution (PMD) and the
PES by solving numerically the TDSE and perform a similar analysis
as it has been done in Refs.
\cite{wollenhaupt,krug,schuricke,JJ,morishita,schuricke2,hart2016,pccp,wessels}.
In order to make a deeper insight into the ionization process, in
addition, we perform a partial-wave analysis of the calculated
PMD. In the next section we describe the model and in
Sec.~\ref{sec:scheme} consider the excitation scheme and
ionization channels. In Sec.~\ref{sec:results} we analyze the
calculated photoelectron momentum distribution and energy spectra.
A summary and conclusions are given in Sec.~\ref{sec:conc}.

\section{The model}
\label{sec:model}

Singly-excited states and the single ionization of the
alkali-metal atoms are, for most purposes, described in a
satisfactory manner using one-electron models. This follows from
the structure of these atoms, which is that of a single valence
electron moving in an orbital outside a core consisting of closed
shells. In that case the valence electron is weakly bound and can
be considered as moving in an effective core potential
$V_\mathrm{core}(r)$, which at large distances $r$ approaches the
Coulomb potential $-1/r$. One of the simplest models for the
effective core potential, applicable for the alkali-metal atoms,
is the Hellmann pseudopotential \cite{hellmann} which reads (in
atomic units)
\begin{equation}
V_\mathrm{core}(r) = - \frac{1}{r} + \frac{A}{r}\,e^{-ar}.
\label{hellmannECP}
\end{equation}
The parameters $A = 21$ and $a = 2.54920$ \cite{MS} provide the
correct value for the ionization potential of sodium $I_\mathrm{p}
= 5.1391\,\mathrm{eV} = 0.18886$\,a.u. and reproduce approximately
the energies of singly-excited states \cite{sansonetti}
(deviations are less than 1\%). The associated eigenfunctions are
one-electron approximations of these states and have the form
$\psi_{nlm}(\mathbf{r}) = R_{nl}(r) Y_{lm}(\Omega)$. Radial
functions $R_{nl}(r)$ can be determined numerically by solving the
corresponding radial equation.

Here we use this single-active-electron approximation to study the
single-electron excitations and ionization of the sodium atom in a
strong laser field. Assuming that the field effects on the core
electrons can be neglected (the so-called frozen-core
approximation \cite{MS}), the Hamiltonian describing the dynamics
of valence (active) electron of the sodium atom in an alternating
field, whose electric component is $F(t) \cos\omega t$, reads (in
atomic units)
\begin{equation}
H = -\frac{1}{2}\nabla^2 + V_\mathrm{core}(r) - F(t) z \cos\omega
t. \label{hamiltonian}
\end{equation}

We consider the linearly polarized laser pulse whose amplitude of
the electric field component (field strength) has the form
\begin{equation}
F(t) = F_\mathrm{peak} \sin^2(\pi t/T_\mathrm{p}),\quad 0 < t <
T_\mathrm{p} \label{pulse}
\end{equation}
[otherwise $F(t) = 0$]. Here $\omega$, $F_\mathrm{peak}$ and
$T_\mathrm{p}$ are the frequency of the laser field, the peak
value of $F$ and the pulse duration, respectively. Since the
system is axially symmetric, the magnetic quantum number $m$ of
the active electron is a good quantum number for any field
strength. In the sodium ground state (when $F = 0$) the orbital
and the magnetic quantum number are equal to zero and in our
calculations we set $m = 0$.

The photoionization process is simulated by solving numerically
the TDSE for the active electron wave function
$\psi(\mathbf{r},t)$ (i.e. by calculating its evolution), assuming
that at $t = 0$ the atom is in the ground state represented by the
lowest eigenstate of Hamiltonian (\ref{hamiltonian}) for $F = 0$.
We have used the second-order-difference (SOD) scheme \cite{askar}
that is for this purpose adapted to cylindrical coordinates
$(\rho,\varphi,z)$ \cite{pccp,epjd2017}. Due to the axial symmetry
of the system, Hamiltonian (\ref{hamiltonian}) and the electron's
wave function do not depend on the azimuthal angle and the
dynamics reduces to two degrees of freedom ($\rho$ and $z$). The
calculations were performed on $1000 \times 2000$ grid in the
wave-packet propagation domain $\rho \le 500$\,a.u.,
$-500\,\mathrm{a.u.} \le z \le 500\,\mathrm{a.u.}$

\section{Energy scheme and photoionization channels}
\label{sec:scheme}

\begin{figure}[!]
\begin{center}
\epsfxsize 0.45\textwidth \epsffile{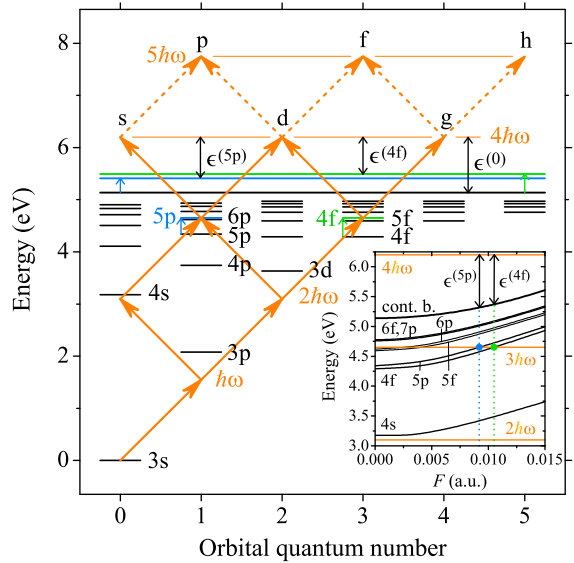}
\end{center}
\vspace{-.2in}
\caption{Energy scheme showing the energy levels (short black
lines) corresponding to singly excited states of sodium
\cite{sansonetti} relative to its ground state (3s) and possible
four-photon and five-photon absorption pathways (orange arrows)
from the ground state to continuum for the radiation of 800\,nm
wavelength ($\hbar\omega \approx 1.55\,\mathrm{eV}$). Levels 5p
and 4f, which are at field strengths $F = 0.0092$\,a.u. and
0.0105\,a.u. shifted into the three photon resonance (see the
inset), are represented by the blue and green short lines (dots in
the inset), respectively. The continuum boundary at these two
values of $F$ and in the weak field limit is in the main plot
represented by horizontal blue, green and black lines,
respectively.} \label{fig:diagram}
\end{figure}

Fig.~\ref{fig:diagram} shows the lowest energy levels
corresponding to singly-excited states of sodium and possible
multiphoton absorption pathways during the interaction of the atom
with a laser radiation of 800\,nm wavelength ($\hbar\omega =
0.05695\,\mathrm{a.u.} \approx 1.55\,\mathrm{eV}$). At this
wavelenghth there are two dominant REMPI channels: (i)
(3+1)-photon ionization via excitation of 5p, 6p and 7p states
(including 2+1+1 process via nearly resonant two-photon transition
$3\mathrm{s} \to 4\mathrm{s}$ and subsequent excitation of
P-states), giving rise to photoelectrons with s and d-symmetry,
and (ii) (3+1)-photon ionization via excitation of 4f, 5f and 6f
states, giving rise to photoelectrons with d and g-symmetry
\cite{krug,hart2016}.

Theoretically, if the multiphoton ionization occurs by absorbing
$N$ photons, the excess energy of ejected electrons in the weak
field limit is $\epsilon^{(0)} = N\!\hbar\omega - I_p$. At
stronger fields, however, the dynamic Stark shift of the ground
state ($\delta E_\mathrm{gr}$), as well as that of the continuum
boundary ($\delta E_\mathrm{cb}$), change effectively the
ionization potential $I_p$ to $I_p - \delta E_\mathrm{gr} + \delta
E_\mathrm{cb}$ and the excess energy becomes dependent on the
field strength (see the inset in Fig.~\ref{fig:diagram}). Within
quadratic approximation $\delta E \approx -\alpha(\omega) F^2\!/4$
one has \cite{pccp}
\begin{equation}
\epsilon(F) \approx N\!\hbar\omega - I_p -
\bigg(\!\alpha_\mathrm{gr}^\mathrm{stat} +
\frac{e^2}{m_{\!e\,}\omega^2}\!\bigg)\frac{F^2}{4},
\label{excess-e}
\end{equation}
where the dynamic polarizability $\alpha(\omega)$ in the ground
state and at the continuum boundary is approximated by its static
value for the sodium ground state
$\alpha_\mathrm{gr}^\mathrm{stat} = 162.7$\,a.u. \cite{mitroy} and
by its asymptotic value in the high frequency limit
$\alpha_\mathrm{cb}(\omega) \approx -e^2/(m_{\!e\,}\omega^2)$,
respectively. Thus, $\delta E_\mathrm{cb} \approx U_p$, where
$U_p$ is the ponderomotive potential of the active electron,
whereas $\delta E_\mathrm{gr} \approx -0.53\,U_p$.

Formula (\ref{excess-e}) for $N = N_0$ (here $N_0 = 4$) and $F =
F_\mathrm{peak}$ gives the energy of photoelectrons whose
contribution in the total yield is maximal, i.e. the position of
main nonresonant peak in the PES. However, if we enter the field
strength values $F_{nl}$ at which the atomic levels $nl$ shift
into resonance with the laser field, the same formula estimates
the positions of REMPI peaks $\epsilon^{(nl)}$ in the spectrum.
The field strengths at which 4f, 5p, 5f and 6p states shift into
the three-photon resonance with the laser field of 800 nm
wavelength are determined in a previous work \cite{pccp}. They are
given in Table~\ref{table1} together with the corresponding values
for $\epsilon^{(nl)}$ obtained by formula (\ref{excess-e}). Note
that the atomic states will be transiently shifted into resonance
twice during the pulse, once as the laser pulse "turns-on" and
again as the pulse "turns-off". Of course, the condition for this
is that $F_\mathrm{peak} > F_{nl}$.

Notice that for the 3+1 REMPI via atomic state $nl$ (which energy
is then in the three-photon resonance with the laser filed, i.e.
$E_{nl} = -I_p + \delta E_\mathrm{gr} + 3\hbar\omega$) formula
(\ref{excess-e}) reduces to $\epsilon^{(nl)} = E_{nl} - U_p +
\hbar\omega$. Since the dynamic Stark shift for the high lying
levels takes approximately the same value as that for the
continuum boundary, the photoelectron energy at the 3+1 REMPI via
considered state will be (see Table~\ref{table1})
\begin{equation}
\epsilon^{(nl)} \approx E_{nl}^{(0)} + \hbar\omega, \label{Eenl-approx}
\end{equation}
where $E_{nl}^{(0)}$ is the energy of the state $nl$ for the
field-free atom. The positions of REMPI maxima in the PES are,
therefore, almost independent on the peak intensity of the laser
pulse, in contrast to the position of the nonresonant four-photon
ionization maximum $\epsilon(F_\mathrm{peak})$. Since usually
$\delta E_{nl} + \delta E_\mathrm{gr} > 0$ (at least for P and F
states, see the inset in Fig.~\ref{fig:diagram}), the states which
can be shifted into three-photon resonance are those with
$E_{nl}^{(0)} \le 3\hbar\omega - I_p$. As a consequence the REMPI
maxima are in the spectrum located below the theoretical value for
photoelectron energy in the weak field limit ($\epsilon^{(nl)} \le
\epsilon^{(0)}$).

\begin{table}[h]
\small \caption{Energies $E_{nl}^{(0)}$ of singly excited P and
F-states ($nl$ from 4f to 7p) of the field free sodium atom
\cite{sansonetti}, field strengths $F_{nl}$ at which these states
shift into the three-photon resonance with the laser field of
800\,nm wavelength [$E_{nl}(F_{nl}) - E_{3s}(F_{nl})=
3\hbar\omega$, $\hbar\omega \approx 1.55$\,eV] and the
photoelectron energies $\epsilon^{(nl)}$ at the 3+1 REMPI via
these states [Eq.~(\ref{excess-e})] \cite{pccp}. For comparison,
$\epsilon^{(nl)}$ values obtained by approximate formula
(\ref{Eenl-approx}) are shown in the fifth column.}
  \label{table1}
  \begin{center}
  \begin{tabular*}{0.45\textwidth}{@{\extracolsep{\fill}}ccccc}
    \hline\hline \\[-3ex]
    state ($nl$) & $E_{nl}^{(0)}$\,(eV) & $F_{nl}$\,(a.u.) &
    $\epsilon^{(nl)}$\,(eV) & $E_{nl}^{(0)} \!+\! \hbar\omega$\,(eV)
    \\[.2ex]
    \hline
    \\[-3ex]
    4f & -0.851 & 0.0105 & 0.707 & 0.699 \\
    5p & -0.795 & 0.0092 & 0.789 & 0.755 \\
    5f & -0.545 & 0.0043 & 1.001 & 1.005 \\
    6p & -0.515 & 0.0028 & 1.035 & 1.035 \\
    6f & -0.378 & - & - & 1.172 \\
    7p & -0.361 & - & - & 1.189 \\
    \hline\hline
  \end{tabular*}
\end{center}
\end{table}

\section{Results}\label{sec:results}
\subsection{Photoelectron momentum distribution}\label{sec:pmd}

The photoelectron momentum distribution (PMD) is determined from
the electron probability density in the momentum space
$|\bar\psi(\mathbf{k},t)|^2$ at $t = T_p$. Transformation of the
wave function from the coordinate to momentum representation can
be done by the Fourier transform. In our case, due to the axial
symmetry of the problem, it is not necessary to calculate the full
3D Fourier transform. The PMD in the $(k_\rho,k_z)$-subspace has
been obtained directly from the outgoing wave part of the function
$\psi(\rho,z)$ by transformation
\begin{equation}
\bar\psi(k_\rho,k_z) = \frac{1}{(2\pi)^2} \int_{-\infty}^\infty
\!\!dz\,e^{-ik_z z} \int_0^\infty \!\! \rho\, d\rho\, J_0(k_\rho
\rho) \psi(\rho,z). \label{ftpsi}
\end{equation}
In order to get a clear PMD, before the transformation one has to
remove the atomic (bound) part of the active electron wave
function $\psi(\mathbf{r},t)$ and leave only the outgoing wave. It
is found that at $t = T_p$ two parts of $\psi(\mathbf{r},t)$
separate approximately at $r = 90$\,a.u.

\begin{figure}
\epsfxsize 0.45\textwidth \epsffile{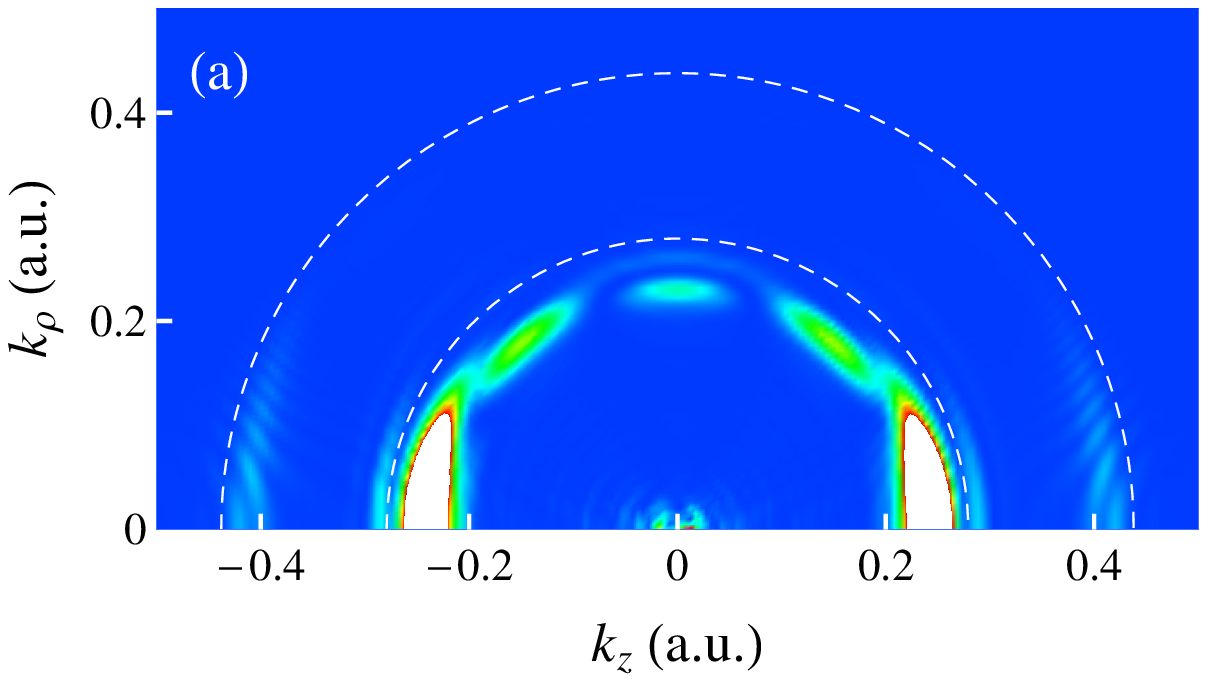}\quad
\\
\epsfxsize 0.45\textwidth \epsffile{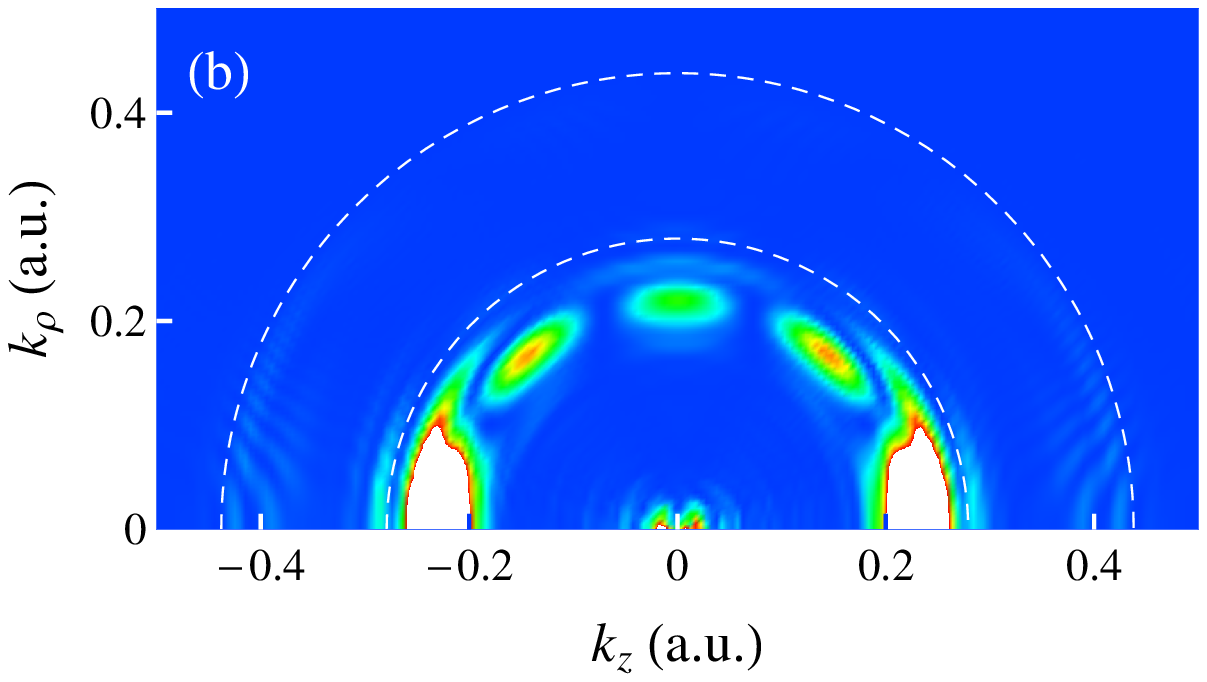}\quad
\\
\epsfxsize 0.45\textwidth \epsffile{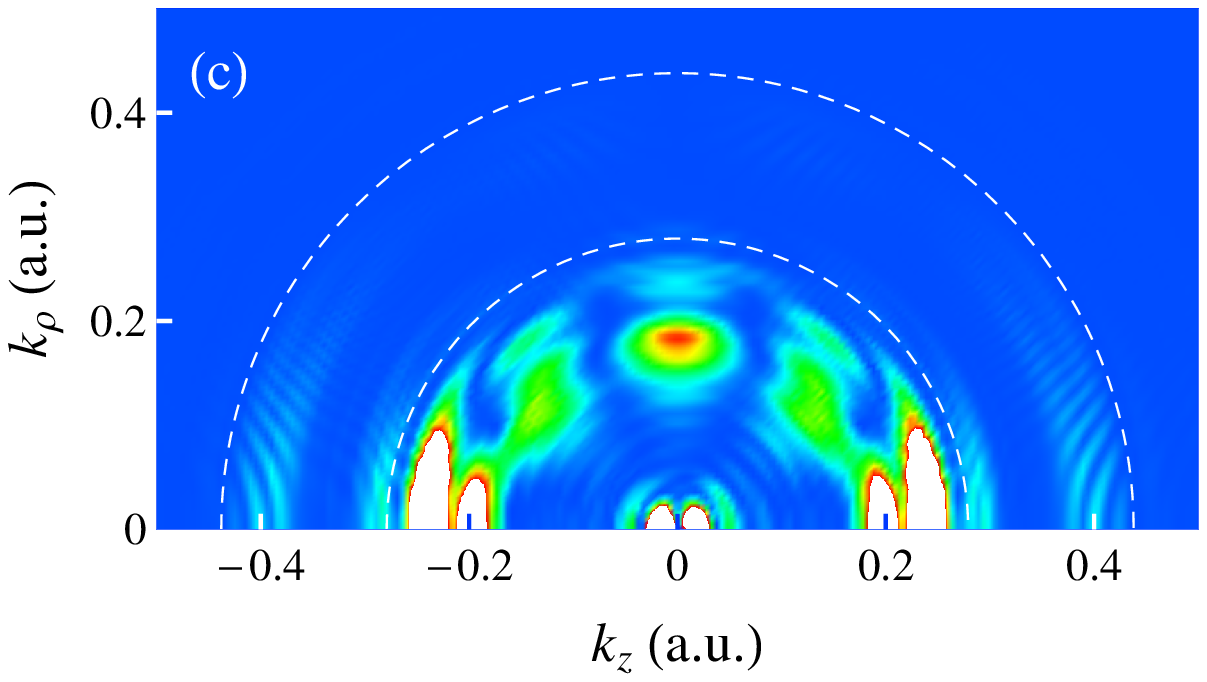}\quad
\caption{(Color online) Photoelectron momentum distribution
$|\bar\psi(\mathbf{k},t)|^2$ in the photoionization of sodium by
the laser pulse ($\lambda = 800$\,nm, $T_p = 57$\,fs) of the form
(\ref{pulse}) calculated at $t = T_p$ for three values of the
laser peak intensity: (a) $3.5$\,TW/cm$^2$, (b) $4.9$\,TW/cm$^2$
and (c) $8.8$\,TW/cm$^2$. The dashed semicircles of radii $k_0 =
0.279$\,a.u. and $k_0^\prime = 0.438$\,a.u. correspond to the
asymptotic values of the photoelectron momentum in the weak field
limit after absorption of four and five photons (the threshold and
the 1st ATI order), respectively.} \label{fig:kdis}
\end{figure}

Fig.~\ref{fig:kdis} shows the calculated PMD for the
photoionization of sodium by 800\,nm wavelength laser pulse of the
form (\ref{pulse}) with 57\,fs duration for three values of the
peak intensity: $3.5$, $4.9$ and $8.8$\,TW/cm$^2$ (the
corresponding field strength are: $F_\mathrm{peak} = 0.0100$,
$0.0118$ and $0.0158$\,a.u.).

The radial ($k$) dependence of the PMD contains information about
the photoelectron energies ($\epsilon = \hbar^2 k^2\!/2m_{\!e}$).
The dashed semicircles of radii $k_0 = 0.279$\,a.u.
($\epsilon^{(0)} = 1.060$\,eV) and $k_0^\prime = 0.438$\,a.u.
($\epsilon^{(0)\prime} = 2.610$\,eV), drawn in the PMD plots, mark
the asymptotic values of momenta (energies) of the photoelectrons
generated in the nonresonant MPI with four and five photons,
respectively, in the weak field limit. Compared to these values,
the radial maxima of PMD determined numerically are shifted toward
the origin of $(k_\rho,k_z)$-plane. (The related energy maxima are
shifted to lower energies, see Sec.~\ref{sec:pwe-pes}.) We point
out that some of these maxima are related to the nonresonant MPI
for different numbers of absorbed photons, while others can be
attributed to the REMPI (Freeman resonances). The shift of
nonresonant maxima $\delta k = \hbar^{-1}
\sqrt{2m_{\!e}\epsilon(F_\mathrm{peak})} - k_0$, referring to
Eq.~(\ref{excess-e}), is determined by the dynamic Stark shift of
the ground state and the continuum boundary at the given laser
peak intensity. The positions of Freeman resonances are, on the
other hand, almost independent on the field strength, but they are
also located below $k_0$ due to inequality $\epsilon^{(nl)} \le
\epsilon^{(0)}$ discussed at the end of Sec.~\ref{sec:scheme}.

The angular structure of the PMD, the so-called photoelectron
angular distribution (PAD), carries information about the
superposition of accessible emitted partial waves, which,
according to selection rules for the four-photon absorption, can
be s, d and g-waves (see Fig.~\ref{fig:diagram}). Indeed, apart
from the strong emission along the laser polarization direction
($\vartheta = 0^\circ$ and $180^\circ$), which can be attributed
to all three partial waves, the PADs also show maxima at
$\vartheta = 90^\circ$, which characterize d and g-waves and at
$\vartheta \approx 45^\circ$ and $135^\circ$, which characterize
the g-wave. Analogously, accessible emitted partial waves for the
five-photon absorption can be p, f and h-waves (see
Fig.~\ref{fig:diagram}).

\subsection{Partial wave expansion of the outgoing wave and
photoelectron energy spectra}\label{sec:pwe-pes}

Generally, the expansion of the outgoing wave in momentum
representation in terms of partial waves reads
\begin{equation}
\bar\psi(\mathbf{k}) = \sum_l \Phi_l(k) Y_{l0}(\vartheta),
\label{superpos}
\end{equation}
where $Y_{l0}(\vartheta)$ are the spherical harmonics with $m = 0$
and $\Phi_l(k) = \int Y_{l0}^*(\vartheta)\, \bar\psi(\mathbf{k})\,
\mathrm{d}\Omega$ are the corresponding radial functions. Using
the representation of $\bar\psi$ in cylindrical coordinates
determined numerically by Eq.~(\ref{ftpsi}), the radial functions
can be calculated as
\begin{eqnarray}
\Phi_l(k) = 2\pi\! \int_0^\pi \!
\bar\psi(k\sin\vartheta,k\cos\vartheta)\, Y_{l0}(\vartheta)\,
\sin\vartheta\,\mathrm{d}\vartheta. \label{radfun}
\end{eqnarray}

\begin{figure}[!]
\epsfxsize 0.45\textwidth \epsffile{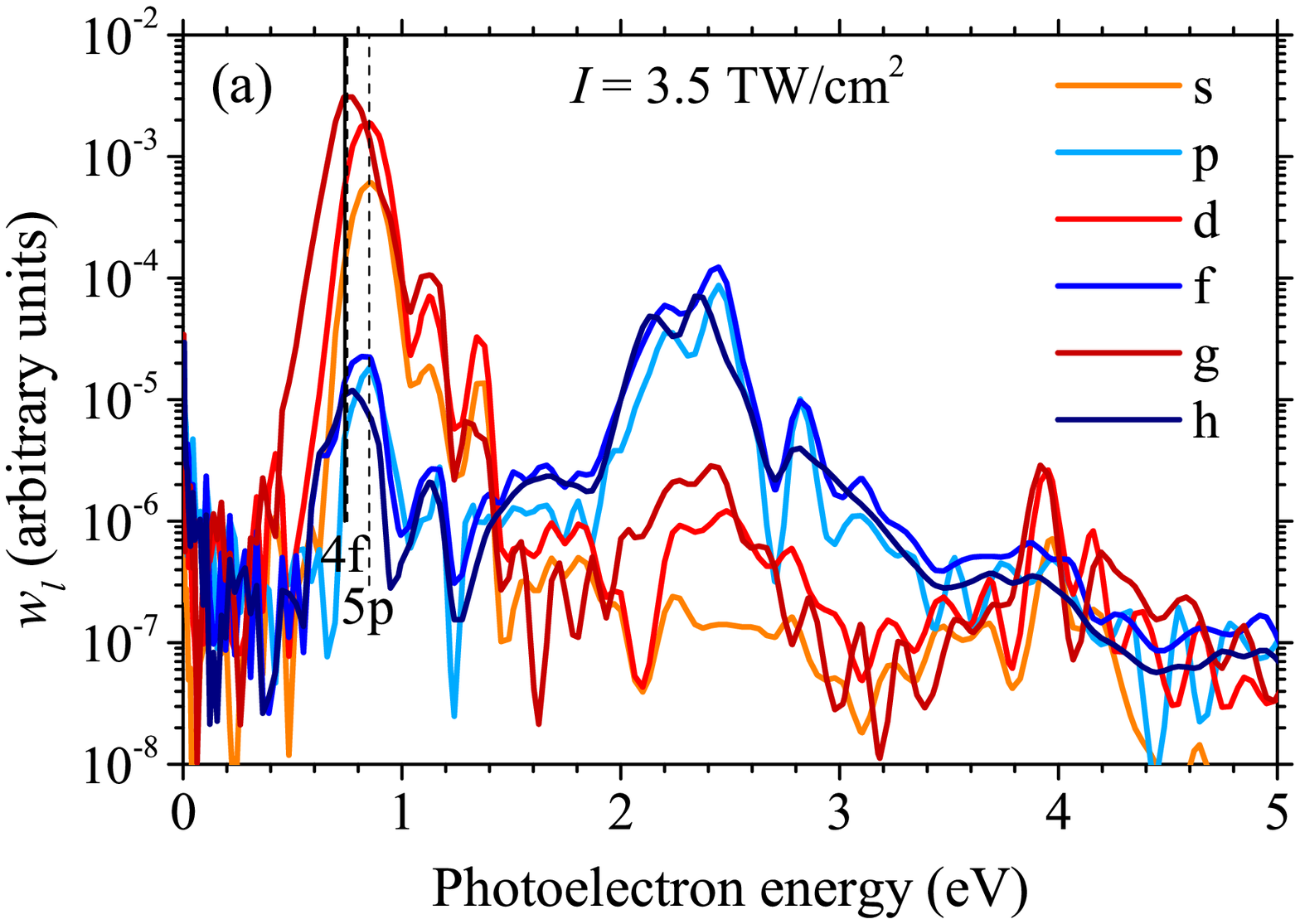}\quad
\\
\epsfxsize 0.45\textwidth \epsffile{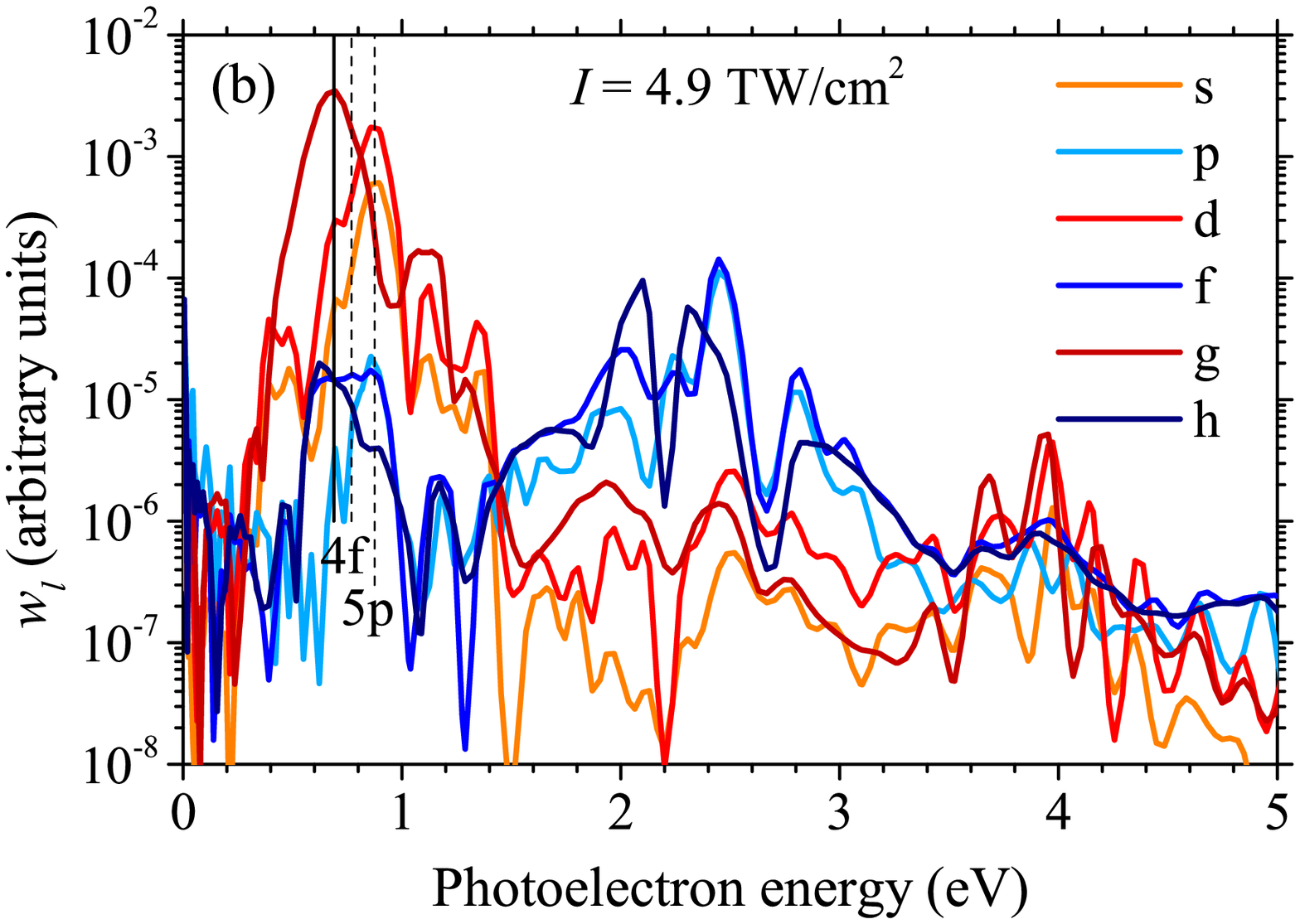}\quad
\\
\epsfxsize 0.45\textwidth \epsffile{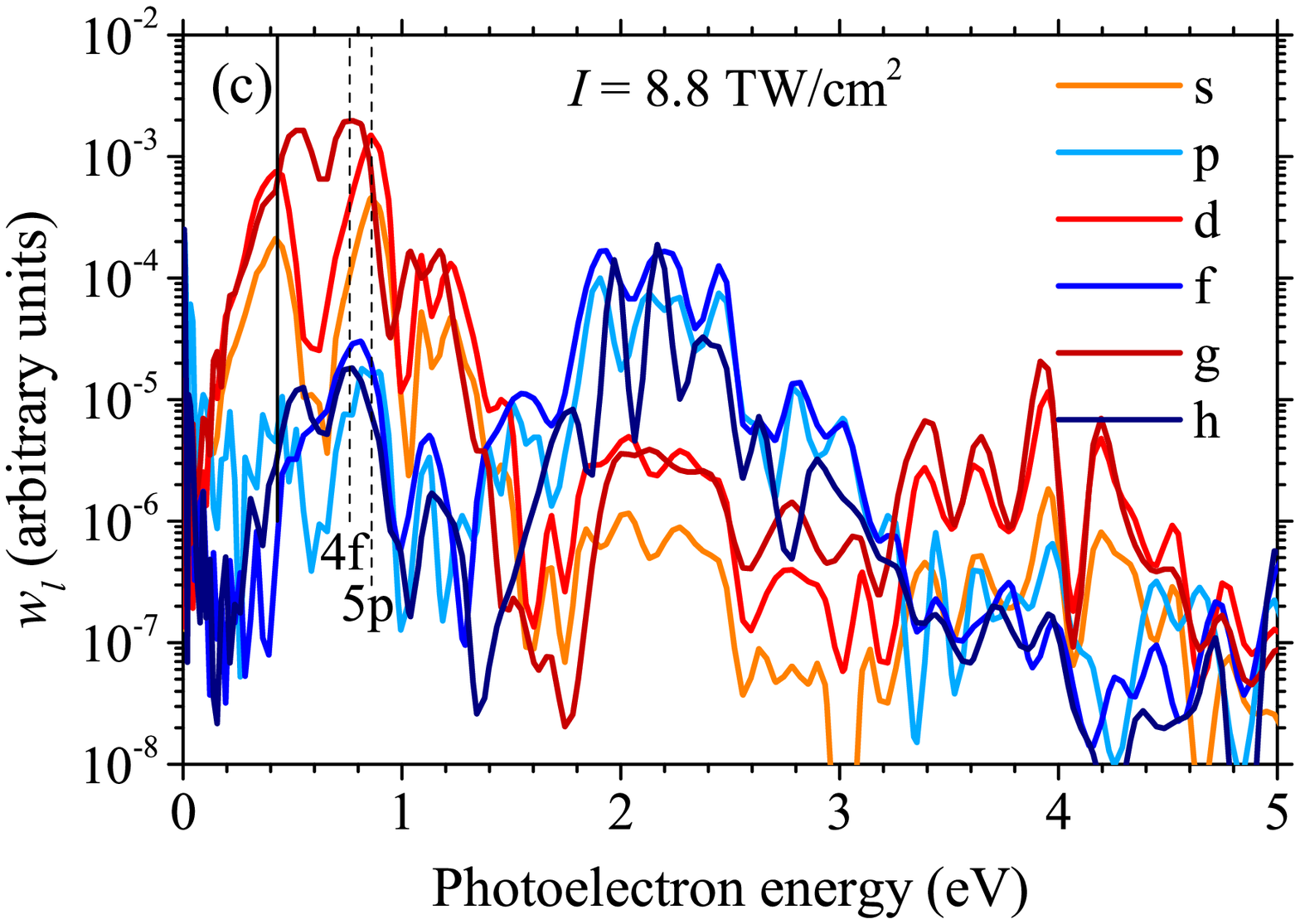}\quad
\caption{(Color online) Partial probability densities
(\ref{partw}) for $l = 0,\ldots,5$ (s, \ldots, h) as functions of
the photoelectron energy $\epsilon = \hbar^2k^2/(2m_e)$, which
correspond to the outgoing waves shown in
Fig.~\ref{fig:kdis}(a)-(c).
The full vertical lines indicate the positions of nonresonant
threshold (four-photon absorption) peak at the corresponding laser
intensities, while the dashed lines mark the energies of two 3+1
REMPI channels (via 4f and 5p state) of the threshold peak.}
\label{fig:wl}
\end{figure}

\begin{figure}
\epsfxsize 0.45\textwidth \epsffile{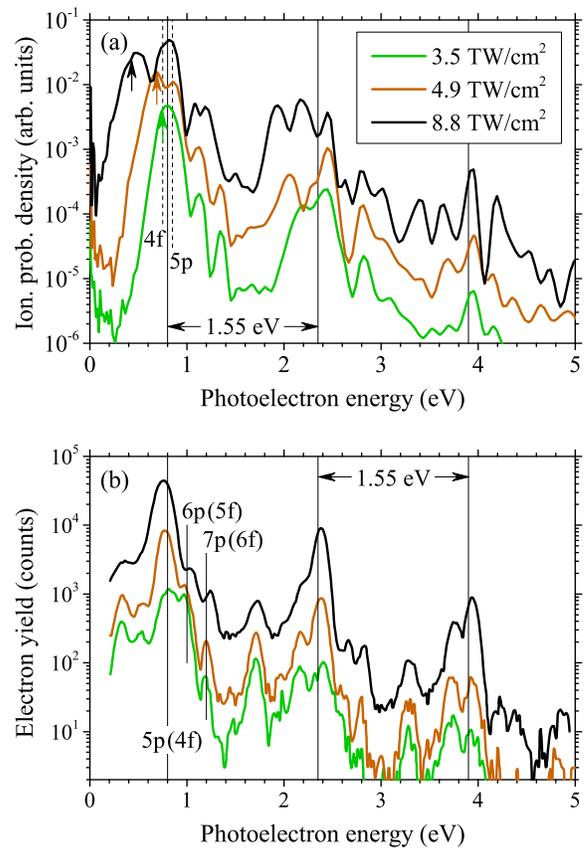}
\caption{(Color online) (a) Total probability density for
photoionization of sodium by the laser pulse with the same
parameters as in Fig.~\ref{fig:kdis} versus the photoelectron
energy, calculated for three values of the laser peak intensity
(3.5, 4.9, 8.8\,TW/cm$^2$). Vertical arrows show the positions of
nonresonant threshold peak at the corresponding intensities, while
the dashed lines mark 3+1 REMPI channels via 4f and 5p state. (b)
Electron yield versus photoelectron energy determined in a recent
experiment \cite{hart2016} for the same pulse parameters as in the
calculations.} \label{fig:pes}
\end{figure}

According to partial wave expansion (\ref{superpos}), the radial
probability density of photoelectrons in momentum space is the sum
$w(k) = \sum_l w_l(k)$, where
\begin{equation}
w_l(k) = |\Phi_l(k)|^2\,k^2 \label{partw}
\end{equation}
are the partial probability densities. These quantities for $l =
0,\ldots,5$, as functions of the photoelectron excess energy
$\epsilon = \hbar^2k^2/2m_{\!e}$, are shown in Fig.~\ref{fig:wl}
for three values of the laser peak intensity: $3.5$, $4.9$ and
$8.8$\,TW/cm$^2$. The corresponding total probability densities
$w$ are shown in Fig.~\ref{fig:pes}(a). The graphs in
Fig.~\ref{fig:pes}(a) represent the photoelectron energy spectra
(PES) for the considered three values of laser intensity. For
comparison, the corresponding spectra obtained experimentally
\cite{hart2016} are shown in Fig.~\ref{fig:pes}(b).

The spectra, both the calculated and experimental, exhibit typical
ATI structure with prominent peaks separated by the photon energy
$\hbar\omega \approx 1.55\,\mathrm{eV}$. Fig.~\ref{fig:pes} shows
the peaks corresponding to lowest three orders of ATI (MPI by
$4+s$ photons, $s = 0,1,2$) which are located approximately at
$\epsilon = 0.8\,\mathrm{eV} + s\hbar\omega$. The partial wave
analysis recovers the character of these peaks. We see in
Fig.~\ref{fig:wl} that for the photoelectron energies around the
threshold peak ($s = 0$, $\epsilon \approx 0.8$\,eV) and around
the second-order ATI peak ($s = 2$, $\epsilon \approx 3.9$\,eV)
dominant contributions in the total probability density come from
the partial waves with even $l$ (s, d, g-waves). Thus, the
photoelectrons with these energies are generated by absorbing an
even number of photons ($N = 4$ and 6). Contrarily, in the
vicinity of the first-order ATI peak ($s = 1$, $\epsilon \approx
2.35$\,eV) the partial waves with even $l$ are suppressed and
those with odd $l$ (p, f, h-waves) dominate. Therefore, in this
case odd number of photons is absorbed (here $N = 5$).

Each ATI peak, in addition, has an internal structure in the form
of local peaks which can be attributed to the nonresonant MPI and
to the REMPI via different excited states.

\subsection{Nonresonant photoionization}\label{sec:nores}

The position of the nonresonant threshold peak (four-photon
ionization maximum) predicted by formula (\ref{excess-e}) for
laser peak intensities 3.5, 4.9 and 8.8\,TW/cm$^2$ is
$\epsilon(F_\mathrm{peak}) = 0.74$\,eV, 0.61\,eV and 0.26\,eV,
respectively. This peak can be observed in Figs.~\ref{fig:wl} and
\ref{fig:pes}(a). Since the energy of photoelectrons produced by
the nonresonant MPI does not depend on $l$, a feature of the
nonresonat peak is that the maxima of contributing partial
densities $w_l$ have the same positions on the energy axis. At the
laser peak intensity of $3.5$\,TW/cm$^2$, however, the nonresonat
peak overlaps with the most prominent REMPI peak [see
Figs.~\ref{fig:wl}(a) and \ref{fig:pes}(a)] and it is difficult to
estimate the position of former from the numerical data. The
position of this peak at intensities $4.9$\,TW/cm$^2$ and
$8.8$\,TW/cm$^2$ is 0.69\,eV and 0.43\,eV (numerical values),
respectively [see Figs.~\ref{fig:wl}(b,c) and \ref{fig:pes}(a)]. A
discrepancy between the values obtained by formula
(\ref{excess-e}) and from numerical calculations is attributed to
the approximative character of the former and to the fact that
probability densities $w_l$ are calculated shortly after the end
of the pulse (not in the asymptotic domain). In addition, it
should be mentioned that in experimental spectra the nonresonant
peak is less prominent (almost invisible). This observation is
reported also in an earlier work presenting a comparison between
calculated and experimental data for the photoionization of
lithium \cite{morishita}. Nonresonant peaks of the first and of
the second ATI order can be observed in Fig.~\ref{fig:pes}, too,
at positions which are shifted by one and two-photon energy
relative to the threshold peak at a given laser intensity.

\subsection{Resonantly enhanced multiphoton ionization}\label{sec:rempi}

In contrast to the nonresonant peaks the positions of REMPI peaks
(Freeman resonances), as explained in Sec.~\ref{sec:scheme}, are
almost independent on the laser peak intensity. We saw that
photoelectrons belonging to the threshold peak reach the continuum
along two pathways which involve the 3+1 REMPI via intermediate P
and F states. For the most prominent peak at $\epsilon \approx
0.8$\,eV the corresponding intermediate states are 5p and 4f,
whereas the subpeaks at $\epsilon \approx 1$\,eV and at $\epsilon
\approx 1.2$\,eV [the positions in Fig.~\ref{fig:pes}(b)] are
related to 3+1 REMPI via states 6p and 5f and via states 7p and
6f, respectively. [The corresponding values in
Fig.~\ref{fig:pes}(a) are slightly shifted upwards since the PMD
and probability densities are calculated immediately after the end
of the pulse.] Note, however, that for a pulse of 800\,nm
wavelength the transfer of population from the ground state to
states 7p and 6f is only near resonant ($E_\mathrm{7p},
E_\mathrm{6f} > 3\hbar\omega - I_p$, see the inset in
Fig.~\ref{fig:diagram}) and, strictly speaking, the four-photon
ionization via these states is not 3+1 REMPI (see the last
paragraph of Sec.~\ref{sec:scheme}). In this case formula
(\ref{excess-e}) is not applicable, but the photoelectron energy
can be estimated using relation (\ref{Eenl-approx}).

Here we focus on the threshold peak at 0.8\,eV. Taking into
account the possible ionization pathways (via states 5p and 4f)
the electron outgoing wave in the energy domain of this peak can
be written as the superposition of two wave-packets
\begin{equation}
\bar\psi = \bar\psi^\mathrm{(5p)} + \bar\psi^\mathrm{(4f)},
\label{wp5p+wp4f}
\end{equation}
which, according to diagram in Fig.~\ref{fig:diagram}, have forms
\begin{eqnarray}
\bar\psi^\mathrm{(5p)} \!\!&=&\!\! \Phi_0^\mathrm{(5p)} Y_{00} +
\Phi_2^\mathrm{(5p)} Y_{20}, \label{wp5p}
\\[.5ex]
\bar\psi^\mathrm{(4f)} \!\!&=&\!\! \Phi_2^\mathrm{(4f)} Y_{20} +
\Phi_4^\mathrm{(4f)} Y_{40}. \label{wp4f}
\end{eqnarray}
Since states 5p and 4f are shifted into the three-photon resonance
at different field strengths (see Table~\ref{table1} and
Fig.~\ref{fig:diagram}), wave packets (\ref{wp5p}) and
(\ref{wp4f}) are formed at different phases of the laser pulse and
characterized by different mean energies ($\approx 0.8$\,eV and
0.7\,eV, respectively, referring to Table~\ref{table1}).

Expression (\ref{wp5p+wp4f}) with components (\ref{wp5p}),
(\ref{wp4f}) is compatible with the partial wave expansion of
function $\bar\psi$. As Fig.~\ref{fig:wl} demonstrates, the
outgoing wave in the domain of threshold peak decomposes into s, d
and g-waves
\begin{equation}
\bar\psi = \Phi_0 Y_{00} + \Phi_2 Y_{20} + \Phi_4 Y_{40}.
\label{decomp}
\end{equation}
Radial functions $\Phi_l$ for considered values of the laser
intensity are determined numerically using formula (\ref{radfun}).
Some parameters of these functions are given in
Table~\ref{table2}. The positions of maxima of $|\Phi_l|$ confirm
the existence of two ionization channels with different energies.
Referring to Table~\ref{table2} the photoelectrons with s and
d-symmetry have higher expected energy ($\approx 0.86$\,eV) than
those with g-symmetry (around $0.76$\,eV). (A discrepancy between
the values for $\epsilon^{(nl)}$ in Tables~\ref{table1} and
\ref{table2} originates from the same reasons as explained in
Sec.~\ref{sec:nores}.) Since the maximum of $|\Phi_2|$ is close to
that of $|\Phi_0|$ we conclude that the majority of d-electrons
are generated in the 3+1 REMPI via 5p state, i.e. their
contribution in the wave packet (\ref{wp4f}) is minor
($\Phi_2^\mathrm{(4f)} \approx 0$). Taking into account the
latest, the comparison between expansion (\ref{decomp}) and
expressions (\ref{wp5p+wp4f}), (\ref{wp5p}), (\ref{wp4f}) gives
$\Phi_0^\mathrm{(5p)} = \Phi_0$, $\Phi_2^\mathrm{(5p)} \approx
\Phi_2$ and $\Phi_4^\mathrm{(4f)} = \Phi_4$.

\begin{table}
\small \caption{Photoelectron energies $\epsilon$ at which the
magnitudes of functions $\Phi_l(k)$ ($l = 0,2,4$) in partial wave
expansion (\ref{decomp}) for the outgoing waves shown in
Fig.~\ref{fig:kdis}(a)-(c) take maximal values and the ratios of
these values $|\Phi_l|/|\Phi_0|$. }
  \label{table2}
  \begin{center}
  \begin{tabular*}{0.45\textwidth}{@{\extracolsep{\fill}}cccccc}
    \hline\hline \\[-3ex]
    partial wave: & s ($l = 0$) & \multicolumn{2}{c}{d ($l = 2$)} & \multicolumn{2}{c}{g ($l =
    4$)}
    \\
    \hline
    $I$ (TW/cm$^2$) & $\epsilon$\,(eV)
                    & $\epsilon$\,(eV) & $|\Phi_2|/|\Phi_0|$
                    & $\epsilon$\,(eV) & $|\Phi_4|/|\Phi_0|$
    \\[.2ex]
    \hline \\[-3ex]
    3.5 & 0.85 & 0.84 & 1.77 & 0.75 & 2.42  \\
    4.9 & 0.88 & 0.87 & 1.68 & 0.77 & 2.65  \\
    8.8 & 0.86 & 0.86 & 1.54 & 0.76 & 2.27  \\
    \hline\hline
  \end{tabular*}
\end{center}
\end{table}

A similar analysis indicates that the subpeaks at $\epsilon
\approx 1$\,eV and at $\epsilon \approx 1.2$\,eV should be related
to 3+1 REMPI via states 5f and 6p and to 3+1 or 2+1+1 REMPI via
state 6f and sequence $4\mathrm{s} \to 7\mathrm{p}$, respectively.
Thus, each of them includes contributions of two ionization
channels of different energies (see Table~\ref{table1}), as in the
case of peak at $\epsilon \approx 0.8$\,eV.

\subsection{Selective enhancement of photoionization channels}
\label{sec:select}

\begin{figure}
\epsfxsize 0.45\textwidth \epsffile{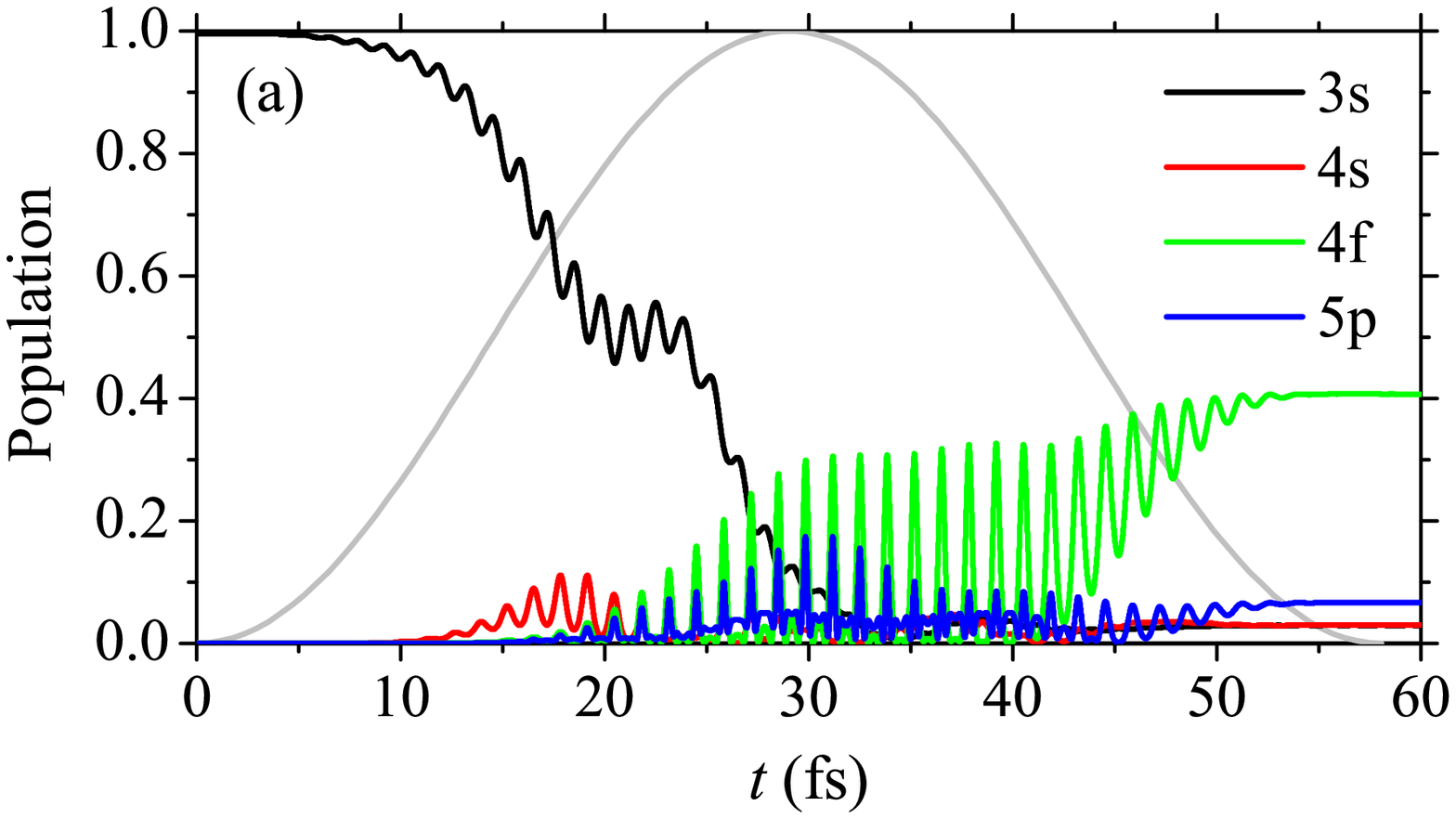}\quad
\\
\epsfxsize 0.45\textwidth \epsffile{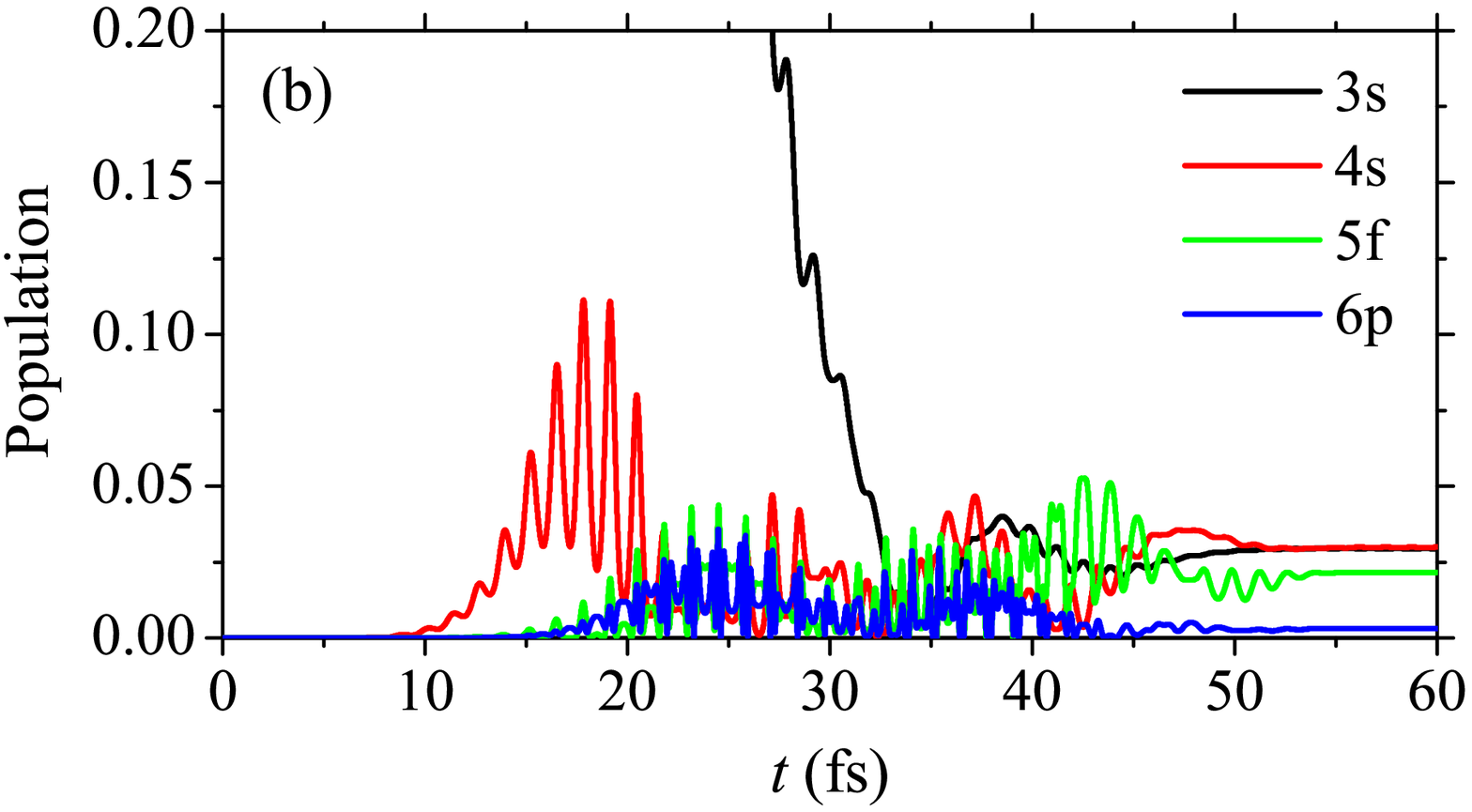}\quad
\\
\epsfxsize 0.45\textwidth \epsffile{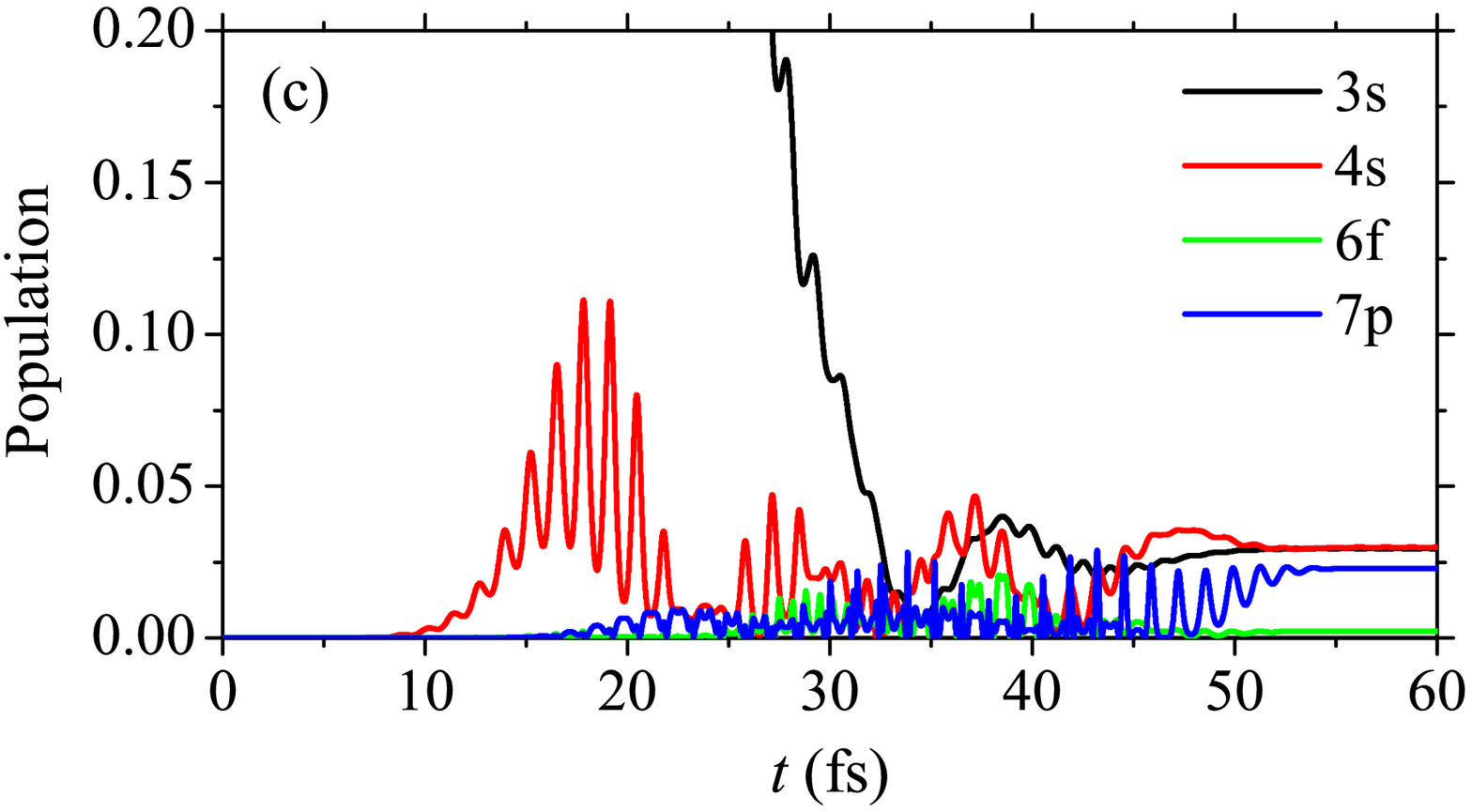}\quad
\caption{(Color online) Populations of the unperturbed ground (3s)
and excited states 4s, 4f, 5p, 5f, 6p, 6f and 7p of sodium during
the 57 fs laser pulse of 3.5\,TW/cm$^2$ peak intensity and 800\,nm
wavelength. The gray line represents normalized envelope
$F(t)/F_\mathrm{peak}$ of the pulse.} \label{fig:pop}
\end{figure}

By comparing the amplitudes of radial functions $|\Phi_l|$ ($l =
0,2,4$) given in Table~\ref{table2} (or alternatively the
corresponding partial densities $w_l$ shown in Fig.~\ref{fig:wl})
it follows that the contribution of g-electrons in the peak at
0.8\,eV is larger than the contributions of d and s-electrons. The
electrons of g-symmetry also dominate in the peak around 1\,eV,
but the largest contribution in the peak around 1.2\,eV is that of
d-electrons (see Fig.~\ref{fig:wl}). Dominant ionization channel
for the peaks at 0.8\,eV and 1\,eV is, therefore, the 3+1 REMPI
via states 4f and 5f, respectively. For the peak around 1.2\,eV,
however, dominant channel is the 2+1+1 REMPI via nearly resonant
two-photon transition $3\mathrm{s}\to 4\mathrm{s}$ and subsequent
excitation of state 7p. The populations of bound states $nl$ of
the unperturbed atom (i.e. transition probabilities $|\langle
nl|\psi(t)\rangle|^2$), calculated during the laser pulse while
solving the TDSE, justify these statements. Fig.~\ref{fig:pop}
shows the populations of 3s, 4s and several P and F unperturbed
states during 57\,fs pulse of 800\,nm wavelength and
3.5\,TW/cm$^2$ peak intensity. Although this intensity corresponds
to the OBI threshold, still there is a significant population of
unionized atoms at all phases of the pulse. It can be seen that
the population of states 4f and 5f is generally higher than that
of states 5p and 6p, respectively, but the population of state 7p
is higher than the population of state 6f. At higher laser
intensities the atoms enter deeply in the OBI domain and in the
second half of pulse the populations significantly drop down (not
shown here) since the majority of atoms becomes quickly ionized.

The populations of relevant states at different phases of the
laser pulse can be well understood by analyzing the energy diagram
for single-electron excitations and taking into account dynamic
Stark shift of energy levels. From the level diagram shown in
Fig.~\ref{fig:diagram} one sees that three-photon transitions from
the ground state to states 4f and 5p are not resonant with the
radiation of 800\,nm in the weak field limit, but these states
shift into resonance at field strength $F \approx 0.01$\,a.u. (see
Table~\ref{table1}), that is in the middle of laser pulse of
3.5TW/cm$^2$ peak intensity. Fig.~\ref{fig:pop}(a) shows that the
population of states 4f and 5p increases rapidly right around
$T_p/2$. Contrarily, three-photon transitions from the ground
state to states 5f and 6p are near resonant with the radiation of
800\,nm in the weak field limit. These states shift into the true
resonance at small values of the field strength, which are reached
two times during the pulse at its opposite sides, as it is visible
in Fig.~\ref{fig:pop}(b).

In a similar way we can analyze the transfer of the population
from the ground state to P states via intermediate 4s state. Since
the two-photon transition $3\mathrm{s}\to 4\mathrm{s}$ is resonant
with the radiation of 777\,nm wavelength in the weak field limit,
applying the laser pulse of 800\,nm wavelength and 3.5TW/cm$^2$
peak intensity will maximally populate the 4s state at the
beginning of the pulse (around $T_p/4$, see Fig.~\ref{fig:pop}),
when the field is not strong enough to shift the state far from
the resonance (see the inset in Fig.~\ref{fig:diagram}). On the
other hand, the single-photon transitions $4\mathrm{s}\to
5\mathrm{p}$, $4\mathrm{s}\to 6\mathrm{p}$ and $4\mathrm{s}\to
7\mathrm{p}$ are in the weak field limit resonant with radiations
of 1075\,nm, 865\,nm and 781\,nm wavelength, respectively.
Therefore, only the transition $3\mathrm{s} \to 4\mathrm{s} \to
7\mathrm{p}$ is fully near resonant and has a significant rate.
Since the dynamic Stark shift for P states increases with the
field strength approximately with the same rate as for the 4s
state (see the inset in Fig.~\ref{fig:diagram}), the transition
$4\mathrm{s}\to 7\mathrm{p}$ remains nearly resonant all the time
and this part of transfer occurs during the rest of the pulse (see
Fig.~\ref{fig:pop}(c)).

We conclude this consideration with a speculation that for the
laser pulse of a shorter wavelength, such that the 4s level during
the pulse transiently shifts into resonance ($\lambda = 780$\,nm
or less), the 2+1+1 REMPI via 4s and subsequent excitation of a
P-state may become more prominent process, increasing in this way
the selectivity of ionization via specific state.

\section{Summary and conclusions}
\label{sec:conc}

In this paper we studied the photoionization of sodium by laser
pulses of 800\,nm wavelength, 57\,fs duration and 3.5 -
8.8\,TW/cm$^2$ peak intensities. This falls into over-the-barrier
ionization (OBI) domain occurring in the multiphoton ionization
(MPI) regime. Using the single-active-electron approximation we
calculated the photoelectron momentum distributions (PMD) by
numerically solving the time dependent Schr\"odinger equation with
these pulse parameters. The contributions of photoelectrons with
different values of orbital quantum number $l$ in the PMD are
determined by expanding the photoelectron wave function in terms
of partial waves. The corresponding partial probability densities
$w_l$ depend on the photoelectron energy $\epsilon$ and the total
density $\sum_l w_l(\epsilon)$ represents the photoelectron energy
spectrum (PES). The spectra calculated for the above mentioned
pulse parameters agree well with the spectra obtained
experimentally by Hart~{\em et.~al.}~\cite{hart2016}.

Partial wave analysis of the spectral peaks related to Freeman
resonances has shown that each peak has photoelectron
contributions from different ionization channels which are
characterized by different photoelectron energies and different
symmetries of released photoelectron wave-packets. It is found
that the most prominent peak around 0.8\,eV is the overlap of two
Freeman resonances related to resonantly enhanced multiphoton
(3+1) ionization (3+1 REMPI) via the states 4f and 5p, but also
has a contribution from the nonresonant four-photon ionization.
The local peak around 1\,eV is related to 3+1 REMPI via the states
5f and 6p, whereas the dominant ionization channel for the peak
around 1.2\,eV is 2+1+1 REMPI via the near resonant 4s state and
subsequently excited 7p state. These findings are justified by
calculating the populations of excited states during the pulse.
Our analysis indicates that the contribution of specific
ionization channels might be selectively increased using laser
pulses of a shorter wavelength, at which the intermediate states
are taken in a better resonance with the laser field.

\bigskip

N.\,S.\,S. thanks J-M. Rost for helpful discussion and gratefully
acknowledges the hospitality at Max-Plank-Institute for the
Physics of Complex Systems in Dresden. We acknowledge support from
the Ministry of Education, Science and Technological Development
of Republic of Serbia under Project No. 171020.

\end{document}